\def\b{\bibitem}
\documentstyle[aps,prl,psfig]{revtex} 
\draft
\begin{document}
\def\SNG{{\em Physical Review Style and Notation Guide}}
\def\LUG {{\em \LaTeX{} User's Guide \& Reference Manual}}
\def\btt#1{{\tt$\backslash$\string#1}}%
\def\REVTeX{REV\TeX}
\def\AmS{{\protect\the\textfont2
        A\kern-.1667em\lower.5ex\hbox{M}\kern-.125emS}}
\def\AmSLaTeX{\AmS-\LaTeX}
\def\BibTeX{\rm B{\sc ib}\TeX}
\twocolumn[\hsize\textwidth\columnwidth\hsize\csname@twocolumnfalse%
\endcsname
\title{Metal-superconductor transition at zero temperature:\\
                 A case of unusual scaling\\
\small{$[$ Phys. Rev. Lett. {\bf 79}, 3042 (1997)$]$}}
\author{T.R.Kirkpatrick}
\address{Institute for Physical Science and Technology, and Department of 
                                                                   Physics\\
         University of Maryland, College Park, MD 20742}
\author{D.Belitz}
\address{Department of Physics and Materials Science Institute,
         University of Oregon,
         Eugene, OR 97403}
\date{\today}
\maketitle
\begin{abstract}
An effective field theory is derived for the normal metal--to--superconductor
quantum phase transition at $T=0$. The critical behavior is 
determined exactly for all dimensions $d>2$. Although the critical 
exponents $\beta$ and $\nu$ do not exist, the usual scaling relations, 
properly reinterpreted, still hold. A complete scaling description of the 
transition is given, and the physics underlying the unusual critical 
behavior is discussed. Quenched disorder leads to anomalously strong
$T_c$-fluctuations which are shown to explain the experimentally observed 
broadening of the transition in low-$T_c$ thin films.
\end{abstract}
]
There has been much interest lately in quantum phase transitions, which
occur at zero temperature ($T=0$) as a function of some nonthermal
control parameter. The subject of much of this attention have been
magnetic transitions\cite{Hertz}. A characteristic feature of these 
transitions is
that the upper critical dimension $d_c^+$, above which the critical
behavior is governed by a simple Gaussian fixed point (FP), is lower
than that of the corresponding classical or finite temperature transition,
if any. The reason is the coupling of statics and dynamics that is inherent
to quantum statistical mechanics, which leads to an effective increase of
the system's dimension from $d$ to $d + z$, with $z$ the dynamical critical
exponent.

In this Letter we study the transition from a normal metal to a superconductor
at $T=0$. We concentrate on the realistic case of a superconductor with a
nonzero density of elastic scatterers, so that the
normal state conductivity is finite. The transition can then be triggered
by varying either the attractive electron--electron interaction, or the 
disorder. For such systems we find $d_c^+ = 2$. For
$d>2$ the transition is governed by a Gaussian FP with unusual properties.
The resulting critical behavior can be summarized as follows. The
correlation length $\xi$ and the order parameter (OP) $\Psi$ scale like
\begin{mathletters}
\begin{equation}
\xi \sim e^{1/2\vert t\vert}\quad,\quad
   \Psi \sim \frac{\Theta(-t)}{\vert t\vert}\,e^{-1/\vert t\vert}\quad,
\label{eq:1a}
\end{equation}
with $t$ the dimensionless distance from the critical point ($t<0$ in the
disordered phase), and $\Theta$ the step function. The physical energy gap 
$\Delta$, 
as measured in a tunneling experiment, is distinct from the OP and scales like
\begin{equation}
\Delta \sim \vert t\vert\,\Psi \sim e^{-1/\vert t\vert}\quad.
\label{eq:1b}
\end{equation}
The relations (\ref{eq:1a}) imply that although
the usual critical exponents $\nu$ and $\beta$, defined
by $\xi \sim \vert t\vert^{-\nu}$ and $\Psi \sim \vert t\vert^{\beta}$,
do not exist ($\nu = \infty$ and $\beta = \infty$), one can assign
a value to their ratio,
\begin{equation}
\beta/\nu = 2\quad,
\label{eq:1c}
\end{equation}
\end{mathletters}%
in the sense that $\Delta\,\xi^2 \sim {\rm const}$. The 
two-point vertex as a function
of the wavevector ${\bf q}$ and the frequency $\Omega$, measured in suitable
units, asymptotically close to criticality has the structure
\begin{mathletters}
\begin{equation}
\Gamma^{(2)}({\bf q},\Omega) = t + \frac{1}{\ln(1/({\bf q}^2 -i\Omega))}\quad.
\label{eq:2a}
\end{equation}
Besides the behavior of the correlation length given above, one reads off
three more critical exponents from this result: The exponent $\eta$, $\gamma$,
and $z$, defined as 
$\Gamma^{(2)}({\bf q},\Omega = 0) \sim \vert{\bf q}\vert^{2-\eta}$, 
$\Gamma^{(2)}({\bf q},\Omega = 0) \sim t^{\gamma}$, and
$\tau_r \sim \xi^z$, with $\tau_r$ the relaxation time, respectively, are
\begin{equation}
\eta = 2\quad,\quad\gamma = 1\quad,\quad z=2\quad,
\label{eq:2b}
\end{equation}
\end{mathletters}%
where $\eta$ is defined only up to logarithmic corrections\cite{LogsFootnote}
We see that
the usual scaling relation $\eta = 2 - \gamma/\nu$ is obeyed (if one
puts $\nu = \infty$ and ignores logarithmic corrections), while all
hyperscaling relations are violated.

The magnetic penetration depth $\lambda$ and the upper critical field 
$H_{c_2}$ scale like the correlation length and the inverse correlation
length squared, respectively,
\begin{equation}
\lambda \sim \xi\quad,\quad H_{c_2} \sim \xi^{-2}\quad.
\label{eq:3}
\end{equation}
The free energy density obeys the scaling law
\begin{mathletters}
\label{eqs:4}
\begin{equation}
f(t,T) = b^{-4}\,f(t\ln b,Tb^2)\quad,
\label{eq:4a}
\end{equation}
with $b$ an arbitrary length rescaling parameter. This yields in particular
for the specific heat coefficient, $\gamma = \partial^2 f/\partial T^2$,
the homogeneity law
\begin{equation}
\gamma(t,T) = \gamma(t\ln b,Tb^2)\quad.
\label{eq:4b}
\end{equation}
\end{mathletters}%
This implies, among other things, that the specific heat coefficient
approaches a constant as the temperature is lowered to zero at the
critical coupling strength. More detailed calculations show a step 
discontinuity in $\gamma(T=0)$ at $t=0$. An analogous scaling behavior 
is found for the electrical resistivity, which obeys
\begin{equation}
\rho(t,T) = \rho(t\ln b,Tb^2)\quad.
\label{eq:5}
\end{equation}
Here the step discontinuity leads from a finite value to zero.
Finally, we find that the fluctuations of the location of the critical point
are long--range correlated, falling off like $1/r^{2(d-2)}$. This 
modifies\cite{WeinribHalperin}
the usual Harris criterion\cite{Harris} for the correlation length exponent to
$\nu\geq 1/(d-2)$, which is obeyed at our Gaussian FP for all $d>2$.
As a result of this long-rangedness, one finds anomalously strong 
fluctuations in the position of the critical point as one
approaches the quantum FP (although the transition remains sharp). 
For the root--mean square fluctuations of the
critical temperature we obtain
\begin{equation}
\Delta T_c/T_c \sim T_c^{(d-2)/2}\,F\bigl(\ln (T_0/T_c)\bigr)\quad,
\label{eq:6}
\end{equation}
with $T_0$ the characteristic temperature of the excitations that mediate
the superconductivity, and $F(x)$ a function that parametrically depends on
the disorder. This explains the strong broadening of the transition
that is observed in thin superconducting films at low values of $T_c$. Our
result is in semi-quantitative agreement with recent mesurements of the
tunneling conductance $G$ in PbBi films\cite{ChervenakValles}. 
In the parameter regime of this experiment, $F(x)\sim x$ is a reasonable
approximation. Defining $\Delta T_c$
as the width of the region over which $G$ drops from 80\% to 20\% of its
normal state value, and using $d=2$\cite{D=2Footnote} and $T_0=300\,{\rm K}$, 
we find that the data of Ref.\ \onlinecite{ChervenakValles}
agree with Eq.\ (\ref{eq:6}) to within 20\%.

The unusual scaling properties listed above are in sharp contrast to the
{\em thermal} critical behavior of superconductors,
which are in the universality class of a $d$-dimensional XY-model. 
In the remainder of this Letter we derive and discuss these results.
Our starting point is a very general fermionic action that we write in
the form
\begin{equation}
S = S_0 - (K_c/\pi N_F^2) \int dq\ {\bar n}_c(q)\,n_c(q)\quad.
\label{eq:7}
\end{equation}
The explicitly written part of this action is the Cooper channel interaction
term, with $K_c$ the appropriate interaction constant normalized by means
of the free electron density of states per spin at the Fermi level, $N_F$.
For an attractive interaction, $K_c <0$.
$q = ({\bf q},\Omega_n)$ comprises space and (bosonic) Matsubara frequency 
labels, and
$\int dq = \sum_{\bf q} T\sum_{i\Omega_n}$. $n_c(q)$ is the Fourier
transform of the anomalous or Cooper channel density
$n_c(x) = \psi_{\downarrow}(x)\,\psi_{\uparrow}(x)$,
where $x=({\bf x},\tau)$ comprises space and imaginary time labels, and
the $\psi_{\sigma}$ are fermionic fields with spin label $\sigma$. 
${\bar n}_c$ is the adjoint of $n_c$.
The remaining part of the action, $S_0$, which we will refer
to as the reference ensemble, describes interacting electrons in the presence
of quenched disorder, with no bare Cooper channel interaction.
Note that even though $K_c =0$ in the bare reference ensemble,
a nonvanishing (repulsive) Cooper channel interaction is generated in
perturbation theory\cite{R}. $S_0$ thus describes a general system of
disordered interacting electrons, with the only restriction being
that it must not undergo any phase transitions in the parameter 
region we are considering, lest the separation of modes that is
inherent in writing the action in the form of Eq.\ (\ref{eq:7})
break down.

We now use standard procedures\cite{Hertz} to derive an OP or
Landau-Ginzburg-Wilson (LGW) effective field theory. We decouple the
Cooper channel interaction by means of a Hubbard--Stratonovich
transformation. This introduces a complex valued OP field, which we denote 
by $\Psi(x)$, 
that couples linearly to $n_c(x)$. The fermionic degrees of freedom are then
integrated out. This way we obtain the partition function in the form
\begin{mathletters}
\label{eqs:8}
\begin{equation}
Z = e^{-F_0/T} \int D[\Psi]\ e^{-\Phi[\Psi]}\quad.
\label{eq:8a}
\end{equation}
Here $F_0$ is the noncritical part of the free energy, and $\Phi$ is
the LGW functional. It can be written
\begin{eqnarray}
\Phi[\Psi]&=&-K_c \int dq\ \vert\Psi(q)\vert^2 
\nonumber\\
&-& \ln\left\langle
   e^{-K_c \int dq\,\bigl(\Psi^*(q)\,n_c(q) + \Psi(q)\,{\bar n}_c(q)\bigr)}
     \right\rangle_{0}\ ,
\label{eq:8b}
\end{eqnarray}
\end{mathletters}%
where $\langle \ldots \rangle_{0} = \int D[\bar\psi,\psi] \ldots e^{S_0}$ 
denotes an average taken with respect to the reference action $S_0$.

We next perform a Landau expansion or expansion of $\Phi$ in powers of
the OP. Since gauge invariance is not broken in the reference ensemble,
only even powers of $\Psi$ appear. The coefficients in this expansion,
i.e. the vertex functions of the effective field theory, are connected
correlation functions of $n_c$ in the reference ensemble.
In particular, the Gaussian vertex is determined by
the pair propagator or anomalous
density--density correlation function in the reference ensemble. Denoting the
latter by $C(q)$, the Gaussian term in the LGW functional
reads
\begin{mathletters}
\label{eqs:9}
\begin{equation}
\Phi^{(2)}[\Psi] = \int dq\ \Psi^*(q)\,\bigl[1/\vert K_c\vert - C(q)
                                         \bigr]\,\Psi(q)\quad,
\label{eq:9a}
\end{equation}
where we have scaled $\Psi$ with $K_c$. So far we have worked within a
particular realization of the disorder. To perform the disorder average
of the free enery we introduce replicas and use a cumulant expansion.
For a particular copy of the replicated system we then obtain, at the
Gaussian level, Eq.\ (\ref{eq:9a}) with $C(q)$ the
connected propagator
$C(q) = \{\langle n_c(q)\,{\bar n}_c(q)\rangle_0\}$, where $\{\ldots\}$ denotes
the disorder average. $C(q)$ is a complicated correlation
function. However, since the reference ensemble is by construction a Fermi
liquid, the structure of this correlation function is known. Renormalization
group arguments show that the structure of $C$ at low frequencies and long 
wavelengths in the limit $T\rightarrow 0$ is\cite{R}
\begin{equation}
C(q) = \frac{Z}{h}\ 
  \frac{\ln\bigl(\Omega_0/(D{\bf q}^2 + \vert\Omega_n\vert)\bigr)}
     {1 + (\delta k_c/h)\,\ln\bigl(\Omega_0/(D{\bf q}^2 
                                             + \vert\Omega_n\vert)\bigr)}\quad,
\label{eq:9b}
\end{equation}
Here $\Omega_0 = k_B T_0/\hbar$ is a frequency cutoff on the order of the 
Debye frequency (for phonon--mediated superconductivity),
and $\delta k_c$ is the repulsive interaction in the Cooper channel that is
generated within perturbation theory even though the bare $K_c$ vanishes.
$D$ is the diffusion coefficient of the electrons,
$h$ is a frequency renormalization constant whose
bare value is $H = \pi N_F/2$, and $Z$ is the wavefunction renormalization.
All of these parameters characterize the reference ensemble, and it is known
that they provide a complete characterization of $C(q)$\cite{Tc}. Using
Eq.\ (\ref{eq:9b}) in Eq.\ (\ref{eq:9a}) we obtain
\begin{equation}
\Phi^{(2)}[\Psi]=\int dq\ \Psi^*(q)\left[t +
   \frac{1}{\ln\bigl(\Omega_0/(D{\bf q}^2 + \vert\Omega_n\vert)\bigr)}\right]
   \Psi(q)\,,
\label{eq:9c}
\end{equation}
\end{mathletters}%
where $t = \left(\delta k_c - Z\vert K_c\vert\right)/h$ and $\Psi$ has been
scaled appropriately.

Let us briefly consider this result for the Gaussian LGW theory. The Gaussian
two-point vertex has the structure of Eq.\ (\ref{eq:2a}), which implies
Eqs.\ (\ref{eq:2b}). By scaling
$\vert{\bf q}\vert$ with the correlation length $\xi$,
we also obtain the behavior of the latter, Eq.\ (\ref{eq:1a}).
Corrections to the Gaussian vertex are known to be of order
$1/\ln^2\Bigl(\Omega_0/(D{\bf q}^2 + \vert\Omega_n\vert)\Bigr)$\cite{R}.
We now need to study the higher order terms in the
LGW functional to ascertain that they do not change these results, and
to obtain information about the OP and the free energy.

We first consider the coefficient of the $\vert\Psi\vert^4$-term in
the Landau expansion. It is a nonlinear anomalous density susceptibility in
the reference ensemble which we denote by $C^{(4)}$.
Due to the cumulant expansion with respect to the
disorder average there are two different contributions to this coefficient,
$C^{(4)} = C_1^{(4)} + C_2^{(4)}$, where $C_1^{(4)}$ is the disorder average
of the four--point correlation function for a given disorder realization,
while $C_2^{(4)}$ is the disorder average of the two--point function squared.
A calculation shows that both $C_1^{(4)}$ 
and $C_2^{(4)}$ are singular in the limit $q\rightarrow 0$.
Cutting off the singularity by means of a wave number
$\vert{\bf p}\vert$, one finds for the leading contributions
$C_1^{(4)} \sim u_4/\vert{\bf p}\vert^4\,\ln^4\vert{\bf p}\vert$ and
$C_2^{(4)} \sim v_4/\vert{\bf p}\vert^{4-d}$, respectively, with $u_4$ and
$v_4$ finite coefficicents\cite{vFootnote}.
The same
method\cite{CoefficientsFootnote} shows that the most divergent contribution
to the coefficient of the term of 
order $\vert\Psi\vert^{2n}$ diverges like
\begin{equation}
C^{(2n)} \sim \frac{u_{2n}}{\vert{\bf p}\vert^{4(n-1)}\,\ln^{2n}\vert{\bf p}
                                                             \vert}\quad,
\label{eq:10}
\end{equation}
with $u_{2n}$ a finite coefficient.
This implies that the Landau expansion of the cutoff regularized LGW theory
is an expansion in powers of $\Psi/{\bf p}^2\ln(1/\vert{\bf p}\vert)$. The
OP field theory, rather than having a simple LGW form, is thus strongly
nonlocal. This is the technical reason for the unusual scaling behavior.

The functional $\Phi$ can be analyzed by using standard 
techniques\cite{Ma}. We are looking for a FP where the
functional dependence of the 2-point vertex on ${\bf q}$ and $\Omega_n$, 
Eq.\ (\ref{eq:9c}), is not renormalized. This fixes the exponents $\eta$ and
$z$. Defining the scale dimension of the correlation length to be
$[\xi] = -1$, Power counting shows that
the coefficients $u_{2n}$ $(n\geq 2)$ of the non-Gaussian terms
have scale dimensions $[u_{2n}]=(n-1)(2-d)$, and hence are irrelevant 
operators with
respect to the FP for all dimensions $d>2$. $[v_4] = -2(d-2)$ for $2<d<4$,
and the higher cumulants are even more irrelevant. The upper critical dimension
is thus $d_c^+ = 2$, and for $d>2$ the critical behavior obtained
from the Gaussian theory is exact.

The scaling behavior of the OP, and of the free
energy, is determined by the term of $O(\Psi^4)$ which
is a dangerous irrelevant operator\cite{MEF} with respect
to these quantities. For scaling purposes, the
cutoff wavenumber $\vert{\bf p}\vert$ can be replaced by the inverse
correlation length, $\vert{\bf p}\vert \sim \xi^{-1}$. The scaling
behavior $\Psi \sim {\bf p}^2\ln(1/\vert{\bf p}\vert)$ observed above
then immediately leads to the behavior of $\Psi$ given in Eq.\ (\ref{eq:1a}). 
One must notice, however,
that the OP function $\Psi$ is distinct from the physical gap function
$\Delta$. The latter determines the gap in the single-particle excitation
spectrum, and hence scales like the frequency or like $\xi^2$. This is
expressed in Eq.\ (\ref{eq:1b}). We have further corroborated this result
by explicit calcuations.\cite{us}

We next consider the critical behavior of the penetration depth $\lambda$ and
the conductivity $\sigma$. Since we have shown that the mean--field/Gaussian
theory yields the exact critical behavior at $T=0$,
all relations between observables that are derived within BCS theory are
valid. In particular, we have $\lambda \sim 1/\sqrt{\Delta}$, and
$H_{c_2} \sim \Delta$ \cite{Tinkham}, which in conjunction with 
$\Delta \sim \xi^2$
yields Eq.\ (\ref{eq:3}). Similarly, we can determine the scale dimension
of the conductivity or resistivity. The real part of the frequency dependent
conductivity in the superconducting phase has a singular contribution,
which within BCS theory is given by
\begin{equation}
{\rm Re}\,\sigma_s(\Omega) = (\pi^2/2)\,\sigma_n\,\Delta\,
                                                      \delta (\Omega)\quad,
\label{eq:11}
\end{equation}
with $\sigma_n$ the conductivity in the normal state. For scaling purposes, 
$\delta(\Omega) \sim 1/\Omega \sim 1/\Delta$. $\sigma_n$ is determined
entirely by properties of the reference ensemble, and hence it does not
show any critical behavior and its scale dimension is zero. We conclude
that the scale dimension of $\sigma_s$ vanishes.
If we assume that the conductivity has only one scaling part, then the
same is true for the conductivity or resistivity in general, and we
obtain Eq.\ (\ref{eq:5}).

We now turn to the free energy density $f$. Hyperscaling suggests that
$f$ scales like $f \sim T/V \sim \xi^{-(d+2)}$, which leads to a
homogeneity law
\begin{equation}
f(t,T,u_4) = b^{-(d+2)}\,f(t\ln b, T b^2, u_4 b^{2-d})\quad,
\label{eq:12}
\end{equation}
where of the irrelevant operators we have written only $u_4$ explicitly.
$f$ is proportional to $u_4\,\Delta^4 \sim 1/u_4$, and hence the
effective scale dimension of $f$ is $[f] = 4$, which yields
Eq.\ (\ref{eq:4a}). Hyperscaling is violated as usual above
an upper critical dimension, viz. by means of the quartic coefficient
being dangerously irrelevant with respect to the free energy. By
differentiating twice with respect to $T$ one obtains the specific heat
coefficient, Eq.\ (\ref{eq:4b}).

The quartic term whose coefficient is $C_2^{(4)}$
yields corrections to scaling that represent fluctuations
in the position of the critical point: By making
$K_c$ a random variable, and integrating out that randomness, one
obtains a term of that structure. Repeating the arguments of
Ref.\ \onlinecite{WeinribHalperin}, we find that the relative fluctuations
of the position of the critical point decay anomalously slowly, 
$\Delta t/t \sim \xi^{-(d-2)}$. $\delta$-correlated elastic scatterers thus
induce long--ranged correlations of the $t$-fluctuations. 
Translating that into the corresponding
fluctuations of $T_c$ via $T_c = T_0\,\exp(-1/\vert t\vert)$ we obtain
Eq.\ (\ref{eq:6}).

Let us now discuss the physics behind the unusual critical behavior
that we have found. The reason for the nonlocal LGW theory and
the ill--behaved Landau expansion is that in deriving the OP field theory
one integrates out the fermions.
This procedure leads to a local OP theory only if the OP fluctuations
or critical modes are the only soft modes in the system. This is not 
necessarily the case, there may be soft modes that are distinct from
the OP fluctuations, but couple to the latter.
These modes may be due to conservation laws, broken symmetries,
or long--range forces. This caveat holds for all phase transitions, including
thermal ones. However, at $T=0$ the number of soft
modes is larger than at finite $T$, and therefore
a nonlocal OP theory is more likely.
Here the additional soft modes are
particle--hole excitations in the Cooper channel that are soft only at
$T=0$. They are integrated out in deriving the OP functional, and they
lead to both the logarithmic structure of the Gaussian vertex, and to
the divergence of the higher vertices. While this leads to a breakdown
of the LGW theory, it also is the reason why the critical behavior
can be determined exactly despite the ill--behaved nature of the OP
field theory: The additional modes lead to an effective long--range
interaction between the OP fluctuations (one that falls off only
logarithmically at large distances and times). For thermal
phase transitions it is well known that long--range interactions stabilize
mean--field theory\cite{FisherMaNickel}, and the same is true here.
Notice that in the present case the long--range interaction
is self--generated by the system.

The same basic physics as discussed above, namely soft modes other than the
OP fluctuations, also makes the quantum critical behavior of
itinerant ferromagnets nontrivial, and yet exactly soluble\cite{fm}.
There are, however, important differences between the two phase
transitions. In the magnetic case, the additional modes are weaker,
which leads to a Gaussian FP with standard power--law critical
behavior. Furthermore, several competing dynamical or temperature
scales lead to complicated dimensionality dependencies of
critical exponents even though the FP is Gaussian. In the
present case the additional modes are so strong
that they completely dominate the physics. As a result, there is
only one temperature scale, with a scale dimension $[T]=2$, and
the quantum critical behavior of disordered bulk superconductors
is BCS-like. However, strong disorder fluctuations lead to a broad
transition region as $d\rightarrow 2$.

We thank Jim Valles, Jay Chervenak, and Shih-Ying Hsu for sharing their 
unpublished data,
and Steve Girvin for bringing Ref.\ \onlinecite{WeinribHalperin} to our
attention. This work was supported by the NSF under grant numbers DMR-95-10185
and DMR-96-32978.

\end{document}